\documentstyle[preprint,aps]{revtex}
\newcommand{\be}{\begin{equation}}
\newcommand{\ee}{\end{equation}}
\newcommand{\ba}{\begin{eqnarray}}
\newcommand{\ea}{\end{eqnarray}}
\begin{document}
\title{How birds fly together: Long-range order in a two-dimensional
dynamical XY model}
\author{Yuhai Tu$^{1}$ and John Toner$^{1,2}$}
\address{{}$^{1}$IBM T. J. Watson Research Center, P. O. Box 218, Yorktown
Heights, NY 10598\\{}$^{2}$
Department of Physics, University of Oregon,
Eugene, OR 97403-1274\cite{byline}
}

\maketitle

\begin{abstract}
We propose a non-equilibrium continuum
dynamical model for the collective motion of large groups of biological
organisms (e.g., flocks of birds, slime molds, etc.)
Our model becomes
highly non-trivial, and different from the equilibrium model, for
$d<d_c=4$; nonetheless, we are able to determine its scaling exponents
{\it exactly} in $d=2$, and show that, unlike equilibrium systems, our
model exhibits a broken
continuous symmetry even in $d=2$. Our model describes a large
universality
class of microscopic rules, including those recently simulated by Viscek
et. al.
\end{abstract}

\newpage

The dynamics of ``flocking" behavior
among living things, such as birds, slime
molds and bacteria has long been
a mystery. Recently, a number of simple numerical models that exhibit such
behavior
have been studied\cite{Viscek,boids}.
For example, reference\cite{Viscek} considers
a synchronous, discrete time step
rule in which
an individual ``bird" in
a group of ``birds" determines its next
direction of motion on each time step by averaging the
direction of its neighbors in certain area, and then adding some
zero mean noise,
while
keeping the magnitude of its velocity
constant. Their simulations in two dimensions find
a transition between an ordered phase in which the mean velocity
of the flock
$<\vec{v} > \neq 0$  and
a disordered phase with $<\vec{v}> = 0$ as the strength of the noise is
increased.

The above two dimensional model
is very similar to the 2D XY model\cite{vpriv,xy}
because the velocity of the ``bird", like the local spin of the
classical XY model, also
has fixed length and continuous rotational symmetry.
Indeed, it is easy to see that, in the limit
that the magnitude of the velocity goes to zero, on each
time step the ``birds" are just picking a new direction, but never
actually move, the model reduces {\it precisely} to the Monte Carlo
dynamics of a two dimensional XY model, with the (small) bird velocity
playing the role of the XY spin.
Since the 2D XY model does {\it not} exhibit a long range ordered phase at
temperatures $T > 0$ (due to spin wave fluctuations),
the long range ordered state observed in reference \cite{Viscek}
seems very surprising. Indeed, in light of the Mermin-Wagner
theorem\cite{mw}
for equilibrium systems, its existence must depend on
fundamentally dynamical, non-equilibrium
aspects of the model.
In this paper, we show, using a
continuum dynamical equation which describes a large universality class
of related dynamical models, that this is indeed the case.
In particular, we explicitly demonstrate:

1) that our model differs from the equilibrium system for spatial
dimensions $d<4$,

2) We can calculate the scaling exponents of this model {\it exactly}
for $d=2$, and

3) the model does, indeed, have a stable spontaneous symmetry
broken state even in two dimensions.

Our starting point is the continuum equations of motion (EOM):
\be
\partial_{t} \vec{v}+(\vec{v}\cdot\nabla)\vec{v}= \alpha \vec{v}-\beta
|\vec{v}|^{2}\vec{v}
-\nabla P
+D_L \nabla (\nabla
\cdot \vec{v})
+D_{1}\nabla^{2}\vec{v}+
D_{2}(\vec{v}\cdot\nabla)^{2}\vec{v}+\vec{f}
\ee
\be
\frac{\partial\rho}{\partial t}+\nabla\cdot(\vec{v}\rho)=0
\ee
where $\beta$, $D_{1}$, $D_{2}$ and $D_{L}$ are all positive, and $\alpha < 0$
in the disordered phase and $\alpha>0$ in the ordered state. The left hand side
of eq. (1) is just
the usual convective derivative of the coarse-grained velocity field $\vec{v}$.
The $\alpha$ and $\beta$ terms simply make the local $\vec{v}$ have a non-zero
magnitude $(=\sqrt{\alpha/\beta})$ in the ordered phase. $D_{L,1,2}$
are diffusion constants.
The Gaussian random noise
$\vec{f}$ has
correlations:
$$
<f_{i}(\vec{r},t)f_{j}(\vec{r}',t')>=\Delta
\delta_{ij}\delta^{d}(\vec{r}-\vec{r}')\delta(t-t')
$$
where $\Delta$ is a constant, and $i$ , $j$ denote Cartesian components.
Finally, the pressure
$$P=P(\rho)=\sum_{n=1}^{\infty} \sigma_n (\rho-\rho_0)^n , $$
where $\rho_0$ is the mean of
the local number density $\rho(\vec{r})$.
The final equation (2) reflects conservation of birds.

The essential difference
between our model and the equilibrium XY model is the existence of the
convective term in our model, which makes the dynamics non-potential
and further stabilizes the ordered phase. A heuristic argument
for the stabilizing effect of the convective term can be given
if we consider our model in Lagrangian coordinates. In those
coordinates, the convective term drops out, and the interaction between
the velocity field is local at each instance. However, at different
times,
the "neighbors" of one particular "bird" will be different depending on
the velocity field itself. Therefore, two originally distant "birds" can
interact with each other at some later time.
It is exactly this
time dependent variable ranged interaction which stabilizes the ordered phase.

To treat the problem analytically, it is more convenient to use the Eulerian
coordinates as in eqs. (1,2). In the rest of our paper, we concentrate on
studying
the symmetry broken phase, where
$\alpha>0$. We
can write the velocity field as $\vec{v}=v_{0}\hat{x}_{||}+\vec{\delta v}$,
where
$v_{0}\hat{x}_{||}=<\vec{v}>$ is the spontaneous average value of $\vec{v}$
We will ignore fluctuations in the magnitude $|\vec{v}|$
from its optimal value of $\sqrt{\alpha/\beta}$ since
they
decay in a finite time (of order $1 \over \alpha$).
Choosing our units of velocity so that $\sqrt{\alpha/\beta}=1$, we can now
write the velocity as:
$
\vec{v}=(\vec{v}_{\perp},\sqrt{1-|\vec{v}_{\perp}|^2})\sim (\vec{v}_{\perp},1-
\frac{1}{2}|\vec{v}_{\perp}|^2)
$,
provided $|\vec{v}_{\perp}|^2<<1$.

Shifting to a co-moving coordinate frame moving
with velocity $\vec {v}_0\hat{x}_{||}$,
$
\vec{v}=(\vec{v}_{\perp},-\frac{1}{2}|\vec{v}_{\perp}|^2)
$, and
the convective term becomes: $(\vec{v}_{\perp}\cdot\nabla_{\perp})\vec{v}_\perp
-
\frac{1}{2}|\vec{v}_\perp|^2\partial_{||}^2\vec{v}_\perp$. We will
neglect the second term,
this will be justified a posteriori. The equation of motion then becomes:

\be
\partial_{t} \vec{v}_{\perp}+\lambda
(\vec{v}_{\perp}\cdot\vec{\nabla}_{\perp})\vec{v}_{\perp}=
-\nabla_{\perp} P+D_L\nabla_\perp(\nabla_\perp\cdot\vec{v}_\perp)
+
D_1\nabla^{2}_{\perp}\vec{v}_{\perp}+D_{||}\partial^{2}_{||}\vec{v}_{\perp}+\vec{f}_{\perp}
\ee
\be
\frac{\partial\delta
\rho}{\partial t}+\rho_0\nabla_\perp\cdot\vec{v}_\perp
+\lambda
\nabla\cdot(\vec{v}\delta\rho)=0
\ee
with $D_{||}=D_{1}+D_{2}$ and $\delta\rho=\rho-\rho_0$.
The
book-keeping coefficient $\lambda = 1$ in the physical case.

We first study the linearized EOM.
We rescale lengths, time, and the fields $\vec{v}_
{\perp}$ and $\delta\rho$
according to
\be
\vec{x}_{\perp}\rightarrow b\vec{x}_{\perp};
x_{||}\rightarrow b^{\zeta}x_{||};
t\rightarrow b^{z}t;
\vec{v}_{\perp}\rightarrow b^{\chi}\vec{v}_{\perp};
\delta\rho\rightarrow b^{\chi_\rho}\delta\rho
\ee
we choose the scaling exponents to keep the diffusion constants
$D_1$, $D_{||}$, $D_\perp =D_1+D_L$ and the strength $\Delta$ of the noise
fixed. The
reason for choosing to keep these particular parameters fixed
rather than, e.g., $\sigma_1$, is that these four parameters completely
determine the size of the equal time fluctuations in the linearized
theory,
as can be seen by solving that theory in Fourier space:
$$
< v^{\perp}_i ({\vec q}, t) v^{\perp}_j (- {\vec q}, t) > =
{\Delta\over 2}
( {q^{\perp}_i q^{\perp}_j\over (D_\perp q_\perp^2 + D_{||} q_{||}^2)
q_\perp^2}+{q_\perp^2 \delta_{ij}^\perp - q^\perp_i q^\perp_j \over
(D_1 q_\perp^2 + D_{||} q_{||}^2)q_\perp^2})\nonumber
$$
The exponents for the linear theory can be determined very easily: $z=2$,
$\zeta=1$, $\chi=1-d/2$, and $\chi_\rho=\chi$ because the
density
fluctuations of $\delta\rho$ are comparable to those of $\vec{v}_\perp$.
Therefore, the linearized theory
implies that $\vec{v}_\perp$ fluctuations grow without bound (like $L^\chi$) as
$L\rightarrow\infty$ for $d\le 2$, where the above expression
for $\chi$ becomes positive. This implies
the loss of long
range order in $d\le 2$.

Making the rescalings
as described above, the other parameters in the model scale as: $\lambda\sim
b^{\gamma_\lambda}\lambda$, $\sigma_n\sim b^{\gamma_n}\sigma_n$
with $ \gamma_\lambda=\chi+1=2-d/2$ and $\gamma_n=z-\chi+n\chi=n+(1-n){d\over
2}$.
The first of these scaling exponents to
become
positive with decreasing $d$ are $\gamma_\lambda$
and $\gamma_2$, which both do so for $d<4$, indicating that
the $\lambda(\vec{v}_\perp\cdot\vec{\nabla})\vec{v}_\perp$ and $\sigma_2
\vec{\nabla}_\perp (\delta\rho^2)$ non-linearities are both
relevant
perturbations for $d<4$. So for $d<4$, the linearized hydrodynamics will
{\it break down}.

An
$\epsilon=4-d$ expansion
will
obviously not be of much use in our problem in $d=2$.
But fortunately, because of the various symmetries
in eqs. (3,4),
we can obtain the {\it exact} scaling exponents
in $d=2$.
First of all, the
reduced equations of motion eqs. (3,4) have a
``Galilean Invariance"\cite{sim}: i.e., if we let:
$\vec{v}_{\perp}(\vec{r},t)\rightarrow\vec{v}_{\perp}(\vec{r},t)+\vec{v}_{\perp,0}
$
and simultaneously boost the coordinate:
$
\vec{x}_{\perp}\rightarrow\vec{x}_{\perp}-\lambda \vec{v}_{\perp,0}t
, $
the equations (3,4) remains invariant for arbitrary values of
$\vec{v}_{\perp,0}$.
This implies that there are no ``graphical" renormalization of the nonlinear
vertex $(\vec{v}_{\perp}\cdot\nabla_{\perp})\vec{v}_{\perp}$, it can only
renormalize by rescaling.
Furthermore, in precisely two dimensions, $D_{||}$ and $\Delta$
are also only renormalized by rescaling. To see this, note that in two
dimensions ${\vec v_\perp}$
has only one component (call it $v_x$),
which can be written as $v_x=\partial_xh$.
The equations of motion (3,4) can then
be rewritten in terms of $h$:
\be
\partial_{t} h+\frac{\lambda}{2} | {\nabla_{\perp}}h |^2=
-\nabla_{\perp}P
+D_{\perp}\nabla^{2}_{\perp}h+D_{||}\partial^{2}_{ ||}h+\eta
\ee
\be
\frac{\partial\delta\rho}{\partial t}+\rho_0 \nabla^2_\perp h + \lambda
\nabla_\perp
\cdot (\delta\rho  \nabla_\perp h) = 0
\ee
where the new effective noise $\eta={{\vec \nabla_\perp}\cdot {\vec f}
\over
\nabla_\perp^2}$ has long ranged correlations. In Fourier space:
$$
< \eta ({\vec q} , \omega) \eta(-\vec{q}',- \omega ')  > = \Delta {\delta^d
({\vec q} -
{\vec q}') \delta (\omega-\omega')\over q_\perp^2} .
$$
This model only corresponds to our original model
(3,4), and hence only describes ``birds", in $d=2$.
However, we will
analyze this model (6,7)
in arbitrary spatial dimensions d,
with the goal of understanding it in the physical case $d=2$.

It is easy
to see that $D_{||}$ and $\Delta
$ cannot be renormalized in this model. $\Delta$ cannot be renormalized
because it is the coefficient of a non-analytic, long-ranged noise-noise
correlation function. The non-linear couplings $\lambda$ and $\sigma_n$
in (6,7), being analytic (i.e., local in space and time), can therefore
{\it not} generate such non-analytic correlations.

The diffusion constant $D_{||}$ likewise cannot be renormalized, because
any such renormalization must clearly involve at least one $\lambda$
with at least one external $h$ leg, and hence at least one power of
$q_\perp$. This can (and
does) renormalize $D_\perp$, but cannot renormalize $D_{||}$.

Implementing the dynamical renormalization group\cite{fut,fns}
we find, quite generally:
\ba
\frac{dD_{||}}{dl}&=&(z-2\zeta)D_{||}\nonumber\\
\frac{d\Delta}{dl}&=&(z+1-d-\zeta-2\chi)\Delta\nonumber\\
\frac{d\lambda}{dl}&=&(\chi+z-1)\lambda\nonumber\\
\frac{dD_{\perp}}{dl}&=&(z-2+G_\perp(\{ g_m \} ))D_{\perp}\nonumber\\
\frac{d\rho_0}{dl}&=&(z-1+G_\rho (\{ g_m \} )\Gamma) \rho_0\nonumber\\
\frac{d\sigma_1}{dl}&=&(z-1+G_1(\{ g_m \} ) \Gamma) \sigma_1\nonumber\\
\frac{d\sigma_n}{dl}&=&(z+(n-1)\chi-1+{G_n(\{ g_m\} )\over g_n}) \sigma_n
\ea
with the parameter $\Gamma={D_\perp\over\sqrt{\sigma_1\rho_0}}$, and the
effective nonlinear coupling constants
$g_1=\frac{\lambda\Delta ^{\frac{1}{2}}}{D_{\perp}^{5/4}D_{||}^{1/4}}$ and
$g_{n\ge 2}=\frac{\sigma_n\Delta^{n-1\over 2} \rho_o^{n \over 2}}
{D_{\perp}^{n+3 \over 4}D_{||}^{n-1 \over 4} \sigma_1^{n \over 2}}$
. The
$G_{\perp,\rho,n}$'s denote the
non-vanishing graphical corrections to $D_\perp$, $\rho_o$, and $\sigma
_n$, respectively. Note that
they explicitly depend {\it only} on the coupling constants $g_m$'s, by
construction.
Note that the absence of graphical corrections to $D_{||}$, $\Delta$,
and $\lambda$
is {\it exact} to all orders in perturbation theory, as discussed
earlier.
Since we seek a fixed point at which $\Delta$, $\lambda$, and
$D_{||}$
remain fixed, we get the following
three exact constraints on the
three exponents:
\be
\chi+z=1;\;\;
z=2\zeta;\;\;
d+\zeta+2\chi=z+1
\ee
whose solution is:
\be
z=\frac{2(d+1)}{5} ; \;\;\;  \zeta=\frac{d+1}{5} ;\;\;\;\chi=\frac{3-2d}{5}
\ee

If we combine the above RG eqs.(8)
, we can obtain the RG flow equations
for the effective coupling constants $g_n$, and the parameter $\Gamma$:
\be
\frac{d\Gamma}{dl}=-(1-G_\perp(\{ g_m\} ))
\Gamma -\frac{1}{2} (G_1(\{ g_m\} ) + G_\rho(\{ g_m\} ))
\Gamma^2
\ee
\be
\frac{dg_1}{dl}=\frac{1}{2}
(4-d-\frac{5}{2}G_\perp(\{ g_m\} ))g_1
\ee

\be
\frac{dg_{n(>1)}}{dl}=\frac{1}{2}
(2n+(1-n)d)g_n+G_n(\{ g_m\} )
+{n\over 2} (G_\rho-G_1) \Gamma -
{n+3\over 4} G_\perp
\ee
from which we see that $g_1$ and $g_2$ become relevant below $d=4$,
while all $g_{n>2}$ are {\it irrelevant} near $d=4$.
Hence we can neglect all $g_{n>2}$'s, and, therefore, all of the
$\sigma_{n>2}$'s as well, at least near d=4.
The parameter $\Gamma$ only becomes relevant below $d=1.5$ where $\chi>0$,
i.e., when  the ordered
phase disappears.
We have calculated
the graphical corrections $G_\perp$ and $G_2$ to one loop order
near $d=4$, and obtain:
$
G_\perp={11 \over 192 \pi^2} ({g_1\over 2} + g_2){g_1\over 2}$,
$G_2=0$.
We do not know whether the vanishing of $G_2$ to this order is the
result of some symmetry of the problem that we have failed to recognize,
in which case $G_2$ would vanish to all orders, or if it is purely
coincidental. In either case, to one loop order, inserting these results
for the graphical corrections into the recursion relations
(16) yields a fixed {\it line} (actually a fixed hyperbola):
$
({g_1\over 2}+g_2)g_1 ={768 \pi^2 \over 55} \epsilon .
$
This summarizes our picture for model (6, 7) in $4-\epsilon$ dimensions.
What happens as we move down
down to two dimensions, which is the only dimension in
which the model (6,7)
actually describes "birds"?
If $G_2$ vanishes to all orders in $\epsilon$, we will still get a fixed
line, as in $4-\epsilon$ dimensions, all the way down to $d=2$, although
its position will be shifted (and it might become curved) away from
that given by the one loop calculation. If $G_2$ does not remain zero, then the
fixed line collapses to a fixed point. In either case, the scaling
exponents continue to be given by equation (10), since those results
depended only on the symmetries of the model.

So in $d=2$, the exponents
are given by equation (10), i.e., $z=\frac{6}{5}$, $\zeta=\frac{3}{5}$,
and $\chi=-\frac{1}{5}$.
These exponents can be checked
experimentally (or from simulations) by measuring, e.g., the
density-density correlation function, which is given, in Fourier space,
by:
\be
< | \rho ({\vec q}, \omega) |^2 > = {\Delta q_\perp^2  \rho_o^2
\over (\omega^2-c^2 q_\perp^2)^2+\omega^2 (D^{R}_{\perp} (\vec{q}, \omega)
q_\perp^2 + D_{||} q_{||}^2)^2}
\ee
where
$c=\sqrt{\sigma_1 \rho_o}$ is the speed of sound, and $D^R_\perp$ is the
renormalized diffusion constant.
As a function of $\omega$, this correlation function (like all
of the correlation and response functions for this
problem)
has two sharp peaks at $\omega=\pm c q_\perp$, of width
$D^{R}_\perp q_\perp^2 + D_{||} q_{||}^2$. Thus, in the frequency
regime containing most of the weight of the correlation function,
$D_\perp^{R}(q_\perp, q_{||}, \omega )$ can be evaluated at
$\omega=c q_\perp$. Using standard renormalization group
arguments and the recursion relation (8)
for $D_\perp$, we find that:
\be
D^{R}_\perp (
{\vec q_\perp}, q_{||}, \omega\sim c q_\perp ; \lambda, \rho_o, \sigma_n)
=q_\perp^{z-2}
f ( {q_{||}\over q_\perp^\zeta})
\ee
Similar RG arguments yield the finite size scaling
of the real-space, real-time rms fluctuations of
$\vec{v}_\perp$:
\be
< | \vec{v}_\perp (\vec {r}, t ) |^2 >
= constant - L_\perp^{2 \chi}
g({L_{||}\over L_\perp^\zeta})
=constant - L_\perp^{-{2\over 5}}
g({L_{||}\over L_\perp^{3\over 5}})
\ee
where $L_\perp$ and $L_{||}$ are the spatial dimensions of the flock
perpendicular to and along the mean direction of motion, respectively,
and in the last equality we have used the value of $\chi$ in $d=2$.
Since this goes to a finite constant as $L\rightarrow \infty$, we see
that long ranged order {\it is stable} in this model in $d=2$, as we
claimed in the introduction.

Now that we have obtained  all the exponents, we need to return to our
original model (1,2) and
verify a posteriori all of our assumptions.
In particular, we must show
that it was
valid to neglect $|\vec{v}_{\perp}|^2\partial_{||}\vec{v}_{\perp}$, which,
under the rescalings eq.(3)
scales like
$\sim b^{2\chi+z-\zeta}\equiv b^{\delta}$. Using the linearized
results for $\chi$, $z$ and $\zeta$,
we
get:
$
\delta=2\chi+z-\zeta=3-d
$
which is clearly less than zero near $d=4$.
Does it remain $< 0$ down to $d=2$? Experience with, e.g., the
$4-\epsilon$
expansion for the $\phi^4$ theory of
critical phenomena suggests that it does. In that problem, a $\phi^6$
perturbation also has {\it linearized}
RG eigenvalue $3-d$. This term nonetheless appears
to remain irrelevant
all the way down to $d=2$, judging by the success of
extrapolations\cite{zinn}
of
$4-\epsilon$ results for the Ising model down to $d=2$. The apparent
contradiction between
this result and the eigenvalue $3-d$, which of course becomes positive
for $d<3$, is that graphical corrections of $O(\epsilon)$ to this result
occur, and keep the eigenvalue negative down to $d=2$. It seems just as
safe to assume that this happens here as in $\phi^4$ theory, and so we
strongly suspect that it does, and that our results for the exponents
do hold {\it
exactly} in $d=2$.

Even in the wildly unlikely event that the cubic vertex {\it does}
become relevant above $d=2$, however,
we can
still show that our model has long-ranged order in $d=2$.
If the cubic vertex does become relevant,
we can no longer obtain the exact scaling exponents in $d=2$, because
both $\lambda$ and $D_{||}$ are now renormalized.

However, not all the scaling relations are lost. The random force is still
unrenormalized since even the contribution from the new vertex
$|\vec{v}_{\perp}|^2
\partial_{||}\vec{v}_{\perp}$ are proportional to $q_{||}^{2}$, which still
vanishes as $|\vec{q}|\rightarrow 0$. Furthermore there is a new scaling
relation coming from the rotational invariance, i. e., since the direction in
which we choose to break the symmetry of $\vec{v}$ was arbitrary, we must,
even after renormalization, be able to resum the nonlinear terms
$(\vec{v}_{\perp}\cdot\nabla)
\vec{v}_{\perp}$ and $|\vec{v}_{\perp}|^2
\partial_{||}\vec{v}_{\perp}$ vertices into the form $(\vec{v}\cdot\nabla)
\vec{v}$. This requires that the graphical corrections to
$(\vec{v}_{\perp}\cdot\nabla)
\vec{v}_{\perp}$ and $|\vec{v}_{\perp}|^2
\partial_{||}\vec{v}_{\perp}$ be the same. To find a fixed point, therefore, we
must have their rescalings to be the same as well. This leads to a new scaling
relation: $2\chi-1=3\chi-\zeta$, or $\chi=\zeta-1$. Taken together with
the last exponents relation in (9), this leads to the following
scaling relation between $z$ and $\chi$:
$
\chi=\frac{z-d}{3} .
$
Now we expect on physical grounds that $z<2$ for all dimensions $d<4$, since
physically, the motion of the ``birds" enhances the mixing, and we know
the corrections due to this effect diverge below $d=4$. Hence we expect
hyperdiffusive behavior, which implies $z<2$. Then the scaling
relation $\chi=\frac{z-d}{3}$ implies $\chi<0$, i.e., true long
range order in $d=2$.

Numerical
simulations\cite{Viscek}
indeed find a long range ordered state in the low
``temperature" regime, in agreement with our predictions above.
Detailed study of the correlation functions to test our predictions for
the scaling exponents (e.g., measurements of
the $\rho-\rho$ correlation function in eqs.(14)) would clearly be of
great interest.

Considerable work remains to be done on this model. The
properties of the low temperature phase of our original model (1,2) in
$d>2$ remain to be determined. Since the symmetries which prevent the
renormalization of the noise strength $\Delta$ and the diffusion
constant $D_{||}$ are lost in $d>2$, it is no longer possible to obtain
exact exponents. However, an $\epsilon$ expansion on the full model (1,2)
should give quite accurate
exponents in $d=3$.

There is also the question of the transition from the ordered to the
disordered state. Without the convective vertex, our model (1,2) is just
model A dynamics for a $\phi^4$ theory, as studied by Halperin et al.
However, we can show that, as in the low temperature phase, the
convective vertex becomes relevant at the transition in $d=4$ as well.
An $\epsilon$ expansion study of this problem is also currently
underway. We will include these subjects and the detailed account of this
letter in future publication \cite{fut}.

We are grateful to T. Viscek for introducing us to this problem,
and providing us with an early draft of
reference 2. We also thank P. Weichman, G. Grinstein, and D. Rokhsar for many
valuable discussions.


\begin{references}
\bibitem[*]{byline}Address after Sept. 16, 1995.
\bibitem{boids} C. Reynolds, {Computer Graphics} {\bf 21}, 25 (1987);
J.L Deneubourg and S. Goss, {Ethology, Ecology, Evolution}
{\bf 1}, 295 (1989); A. Huth and C. Wissel, in {\em Biological
Motion}, eds. W. Alt and E. Hoffmann (Springer Verlag, 1990)p. 577-590.
We thank D. Rokhsar for calling these.
references to our attention.
\bibitem{Viscek} T. Viscek et. al., preprint.
\bibitem{vpriv} This was first pointed out to us by T. Viscek.
\bibitem{xy} J. M. Kosterlitz and D. J. Thouless,
{J. Phys. C} {\bf 6}, 1181 (1973).
\bibitem{mw} N. D. Mermin and H. Wagner, {Phys. Rev. Lett.} {\bf 17},
1133(1966).
\bibitem{sim} This is quite similar to the behavior of the
incompressible Navier-Stokes equation forced at zero wavevector, which
also has a critical dimension $d_c=4$ below which linearized
hydrodynamics breaks down, and for which it is also possible to
obtain exact exponents. See reference \cite{fns}.
\bibitem{zinn} J. C. LeGuillou and J. Zinn-Justin, {J. de Phys. Lett.}
{\bf 46}, L-137 (1985).
\bibitem{fns} D. Forster, D. R. Nelson, and M. J. Stephen,
{Phys. Rev. A} {\bf 16}, 732 (1977).
\bibitem{fut} J. Toner and Y. Tu, in preparation.
\end{references}
\end{document}